# The Impact of IR-based Classifier Configuration on the Performance and the Effort of Method-Level Bug Localization

Chakkrit Tantithamthavorn[a,∗], Surafel Lemma Abebe[b], Ahmed E. Hassan[c], Akinori Ihara[d], Kenichi Matsumoto[d]

[a]*The University of Adelaide, Australia.*
[b]*The Addis Ababa University, Ethiopia.*
[c]*Queen's University, Canada.*
[d]*Nara Institute of Science and Technology, Japan.*

## Abstract

*Context:* IR-based bug localization is a classifier that assists developers in locating buggy source code entities (e.g., files and methods) based on the content of a bug report. Such IR-based classifiers have various parameters that can be configured differently (e.g., the choice of entity representation).
*Objective:* In this paper, we investigate the impact of the choice of the IR-based classifier configuration on the top-k performance and the required effort to examine source code entities before locating a bug at the method level.
*Method:* We execute a large space of classifier configuration, 3,172 in total, on 5,266 bug reports of two software systems, i.e., Eclipse and Mozilla.
*Results:* We find that (1) the choice of classifier configuration impacts the top-k performance from 0.44% to 36% and the required effort from 4,395 to 50,000 LOC; (2) classifier configurations with similar top-k performance might require different efforts; (3) VSM achieves both the best top-k performance and the least required effort for method-level bug localization; (4) the likelihood of randomly picking a configuration that performs within 20% of the best top-k classifier configuration is on average 5.4% and that of the least effort is on average 1%; (5) configurations related to the entity representation of the analyzed data have the most impact on both the top-k performance and the required effort; and (6) the most efficient classifier configuration obtained at the method-level can also be used at the file-level (and vice versa).
*Conclusion:* Our results lead us to conclude that configuration has a large impact on both the top-k performance and the required effort for method-level bug localization, suggesting that the IR-based configuration settings should be carefully selected and the required effort metric should be included in future bug localization studies.

## 1. Introduction

Developers spend 50% of their programming time debugging the source code in an unfamiliar software system [1]. Debugging mainly includes locating buggy source code entities and fixing them. Establishing a good strategy to help developers quickly locate buggy source code entities considerably reduces developers' effort and debugging time. To this end, several studies propose the use of Information Retrieval (IR) based classifiers for bug localization [2, 3, 4, 5, 6, 7, 8, 9, 10, 11, 12, 13, 14, 15].

IR-based classifiers have different configuration parameters. For example, source code entities can be represented using only identifiers or using comments and identifiers. Recent studies suggest that such configuration parameters may impact the top-$k$ performance of IR-based bug localization [16, 17]. In a recent study, Thomas et al. [2] show that the choice of classifier configurations impacts the performance of IR-based classifiers at the file-level granularity. In addition to file-level, IR-based classifiers are often used to locate bugs at the method-level [18, 19, 20, 21, 22, 23, 24]. Indeed, our recent work shows that method-level bug localization requires less effort to locate bugs than file-level bug localization [25]. However, little is known about the impact that the choice of a classifier configuration has on classifiers that are used to locate bugs at the method-level.

In this paper, we partially replicate and extend Thomas et al. [2] to investigate the impact of IR-based classifier configuration on the top-$k$ performance and the required effort to examine source code entities (e.g., files and methods) before locating a bug at the method level. Moreover, we also analyze the classifier sensitivity to parameter value changes. Finally, we investigate whether the most efficient classifier configuration for file-level bug local-

∗Corresponding author.
  *Email addresses:*
  `chakkrit.tantithamthavorn@adelaide.edu.au` (Chakkrit Tantithamthavorn), `surafel.lemma@aait.edu.et` (Surafel Lemma Abebe), `ahmed@cs.queensu.ca` (Ahmed E. Hassan), `akinori-i@is.naist.jp` (Akinori Ihara), `matumoto@is.naist.jp` (Kenichi Matsumoto) ization is also the most efficient at the method-level (and vice versa). In total, we explore a large space of classifier configurations 3,172 configurations. Through a case study of 5,266 bug reports of two software systems (i.e., Eclipse and Mozilla), we address the following research questions:

- **(RQ1) Can IR-based classifier configurations significantly impact the <u>top-$k$ performance</u> of method-level bug localization?**
  The choice of classifier configuration impacts the top-$k$ performance from 0.44% to 36%, indicating that using an inappropriate configuration could result in poor top-$k$ performance. Moreover, there are only few classifier configurations that perform close to the best performing configuration, indicating that finding the best top-$k$ configuration is difficult.

- **(RQ2) Can IR-based classifier configurations significantly impact the <u>required effort</u> for method-level bug localization?**
  The required effort of the classifier configurations vary from 4,395 to 50,000 LOC, indicating that using an inappropriate configuration could result in wasted effort. Classifier configurations which give similar top-$k$ performance often require different amount of effort, suggesting that practitioners should take into consideration the effort that is required to locate bugs instead of simply using the top-$k$ metrics when comparing the performance of classifier configurations.

- **(RQ3) Is the most efficient classifier configuration for method-level bug localization also the most efficient configuration for file-level bug localization (and vice versa)?**
  The most efficient classifier configuration obtained at the method-level can also be used at the file-level (and vice versa) without a significant loss of top-$k_{LOC}$ performance.

Our results lead us to conclude that configuration has a large impact on both the top-$k$ performance and the required effort for method-level bug localization. The results suggest that the IR-based configuration settings should be carefully selected and the required effort metrics should be included in future bug localization studies. Nonetheless, we find that configurations related to the entity representation of the analyzed data have the most impact on the top-$k$ performance and the required effort, suggesting that practitioners would benefit from guidance on which configuration parameters matter the most.

### 1.1. Paper Organization

The remainder of the paper is structured as follows. Section 2 introduces Information Retreival (IR)-based Bug Localization. Section 3 motivates our research questions, while Section 4 describes our case study design. The results of our case studies are presented in Section 5. Threats to the validity of our study are disclosed in Section 6. Finally, Section 7 draws conclusions.

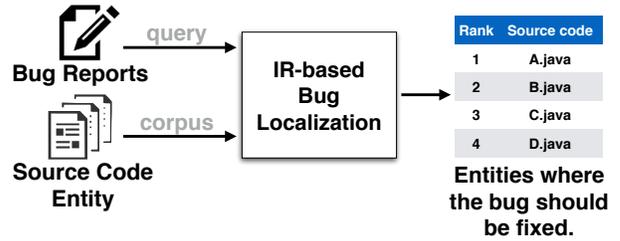

Figure 1: An overview of IR-based Bug Localization.

## 2. IR-based Bug Localization

Bug localization is the task of identifying source code entities that are relevant to a given bug report. In the literature, various approaches that exploit information such as dynamic and textual information are proposed in order to support bug localization. A comprehensive survey of the approaches can be found in Dit *et al.* [26]. In this section, we discuss approaches that exploit textual information using Information Retrieval (IR) classifiers to locate source code entities that are relevant to fix a bug.

In IR-based bug localization (see Figure 1), a bug report is treated as a query and source code entities are treated as a document corpus. A source code entity is considered *relevant* if the entity indeed needs to be modified to resolve the bug report, and *irrelevant* otherwise. A source code entity could be defined at different granularity levels (e.g., package, file, or method). Prior studies use file [5, 6, 2, 7, 27], or method [21, 22, 24, 23, 25] granularity levels.

**File-level bug localization** is well-studied using several popular IR models such as Vector Space Model (VSM) [6], Latent Semantic Indexing (LSI) [24, 23], and Latent Dirichlet Allocation (LDA) [24, 23]. The results of the studies, however, cannot be generalized. For example, Rao and Kak [6] state that VSM outperforms sophisticated classifiers like LDA, while Lukins *et al.* [24, 23] state that LDA outperforms LSI. As noted in Thomas *et al.* [2], the differences are often due to the use of different datasets, different evaluation metrics, and different classifier configurations. The classifier configuration for IR-based bug localization includes the choice of bug report representation, entity representation, preprocessing steps, and IR-classifier parameters. To compare different IR-based classifiers, Thomas *et al.* [2] proposed a framework with a large space of classifier configurations. Thomas *et al.* [2]'s results show that the configuration of IR-based classifiers impacts the performance of the IR-based bug localization at the file level.

**Method-level bug localization** considers a relatively smaller source code entities (i.e., methods) instead of file. Marcus et al. [18] use LSI to identify the methods that are required to implement a task. The results show that LSI provides better results than existing approaches at that time, such as regular expressions and dependency graphs. Marcus et al. [19] and Poshyvanyk et al. [20] combine LSI with a dynamic feature location approach called scenario-based probabilistic ranking in order to locate methods relevant to a given task. Lukins et al. [24, 23] use Latent Dirichlet Allocation (LDA) techniques to localize bugs at the method-level. Their study on Eclipse, Mozilla, and Rhino shows that LDA-based approach is more effective than approaches using LSI. However, to evaluate their approach, they use only 322 bugs across 25 version of three systems (Eclipse, Mozilla, and Rhino), which on average are less than 20 bugs per version. Wang et al. [22] investigate the effectiveness of several IR classifiers on method-level bug localization. The results of Wang et al. [22]'s study show that older and simpler IR techniques, e.g., VSM, outperform more recent IR techniques. In their study, however, they use only one classifier configuration. Hence, their results cannot be validated and generalized to other classifier configurations.

Our RQ1 revisits Thomas et al. [2]'s study at the method level. Moreover, we investigate the difficulty of locating optimal configurations and the impact of each parameter on the overall performance of a classifier. Such sensitivity analysis of parameters provides us with a better understanding of the difficulty of the problem of finding configurations in practice.

**Evaluation Metrics** are used to assess the performance of different classifiers in localizing bugs. Several studies use the top-$k$ performance metric to carry out such assessment [5, 6, 2, 7, 3, 28]. The top-$k$ performance metric considers a bug to be localized if at least one relevant source code entity is returned in the top-$k$ ranked entities. Metrics such as Precision, Recall, Mean Average Precision (MAP), and Mean Reciprocal Ranks (MRR) will not be used for our study, since practitioners are primarily interested in the top-$k$ suggested entities Guan et al. [29].

In contrast to recent studies of effort-aware bug prediction [30, 31, 32], prior studies of bug localization have not explored the effort that is needed to cope with irrelevant suggestions until a relevant entity is located. Hence, our RQ2 explores this notion of effort as another dimension to evaluate and compare classifiers for IR-based bug localization.

## 3. Research Questions

The goal of our study is to better understand: (1) the impact of classifier configurations on method-level bug localization and which parameters have a large impact on the performance of IR-based classifiers; (2) the impact of classifier configurations on the required effort to locate bugs at the method level and which parameters have a large impact on the required effort of IR-based classifiers; and (3) whether the most efficient classifier configuration found for file-level bug localization is also efficient at method-level (and vice versa).

To do so, we executed a large space of classifier configuration, 3,172 in total, on 5,266 bug reports of two software systems, i.e., Eclipse and Mozilla. We define and present the rationale of our research questions below:

> (RQ1) Can IR-based classifier configurations significantly impact the **top-$k$ performance** of method-level bug localization?

**Motivation.** Thomas et al. [2] showed that the choice of classifier configuration impacts the top-$k$ performance of file-level bug localization. However, the impact of classifier configurations on method-level bug localization and a comparison to file-level bug localization remains largely unexplored. Besides, prior research has paid attention to identifying the ideal IR configuration. Lohar et al. [33] use a Genetic Algorithm to identify the best IR configuration for traceability link recovery. Panichella et al. [34] use a Genetic Algorithm to determine the best LDA configuration. Yet, little is known which parameters of IR-based bug localization truly have the most impact on the top-$k$ performance. For example, the choice of preprocessing techniques might have a larger impact than the choice of LDA configurations. Knowing which parameters are influential indicators of the top-$k$ performance could help practitioners make effective use of their time when exploring various parameters (e.g., not spending too much time to find the optimal LDA configurations).

> (RQ2) Can IR-based classifier configurations significantly impact the **required effort** for method-level bug localization?

**Motivation.** Traditionally, the performance of IR-based bug localization approaches is evaluated using the top-$k$ performance metric. However, the top-$k$ performance metric does not take into consideration the effort that is required to examine the entities that are ranked before the first relevant source code entity Xu et al. [35]. The metric assumes that the required effort to locate a relevant entity ranked first and tenth are the same, which is not necessary true. While for a developer, the required effort is different due to the number of ranked entities to be examined and their varying sizes. Yet, little is known about the impact that classifier configuration has on the required effort to locate the first buggy entity. Knowing the required effort provides us a better understanding of the practicality of a classifier [36, 31].

> (RQ3) Is the most efficient classifier configuration for method-level bug localization also the most efficient configuration for file-level bug localization (and vice versa)?

**Motivation.** In RQ2, we investigate if the top-$k$ performer requires the least effort to locate bugs. Therefore, in addition to evaluating the performance and the effort individually, investigating the most efficient configuration is also necessary for practitioners [31, 37]. An *efficient* configuration is a configuration which gives the best top-$k$ performance with a limited amount of reviewing effort. Since traditional IR evaluation metrics (e.g., top-$k$ performance) do not consider the required effort [35], the most efficient configuration still remains unexplored. Besides, different researchers conduct bug localization only at one granularity level, i.e., file or method. Prior research finds that classifiers for bug prediction that are designed to work well at one granularity level do not often work well at another level [30, 38, 39]. Yet, little is known about whether such performance variances hold for IR-based bug localization at different granularity levels. RQ3 investigates if the configuration found to be efficient at a granularity level will also be efficient on the other granularity levels. Knowing the most efficient configuration of bug localization would help practitioners choose the configuration that performs best and requires the least effort.

Table 1: Studied systems.

|  | Eclipse (JDT) | Mozilla (mailnews) |
|---|---|---|
| Domain | IDE | Web browser |
| Language | Java | C/C++/Java |
| Years considered | 2002 - 2009 | 2002 - 2006 |
| # Bug reports | 3,898 | 1,368 |
| Source code snapshots | 16 | 10 |
| # Source code files | 1,882 - 2,559 | 319 - 332 |
| # Source code methods | 17,466 - 27,404 | 6,656 - 7,466 |
| Source code corpus size | 232 - 506 (KLOC) | 173 - 193 (KLOC) |

Table 2: Descriptive statistics of the studied systems.

|  | Min. | $1^{st}$ Qu. | Med | Mean | $3^{rd}Qu.$ | Max |
|---|---|---|---|---|---|---|
| **Eclipse system** | | | | | | |
| Bug report size (#words) | 2 | 40 | 67 | 146 | 129 | 6,187 |
| File size (LOC) | 1 | 59 | 116 | 255 | 255 | 12,490 |
| Method size (LOC) | 0 | 4 | 8 | 14 | 15 | 2,518 |
| # files per bug report | 1 | 1 | 1 | 2 | 2 | 10 |
| # methods per bug report | 1 | 1 | 3 | 15 | 8 | 98 |
| **Mozilla system** | | | | | | |
| Bug report size (#words) | 4 | 54 | 96 | 127 | 152 | 2,077 |
| File size (LOC) | 41 | 162 | 448 | 777 | 885 | 8,733 |
| Method size (LOC) | 0 | 9 | 19 | 35 | 42 | 720 |
| # files per bug report | 1 | 1 | 1 | 2 | 2 | 9 |
| # methods per bug report | 1 | 1 | 3 | 18 | 10 | 96 |

## 4. Case Study Design

In this section, we provide a summary of the studied systems, and our data extraction and analysis approaches.

### 4.1. Studied Systems

In order to address our three research questions and establish the validity and generalizability of our results, we conduct an empirical study using two of the three software systems that are provided by Thomas *et al.* [2] (i.e., Eclipse and Mozilla). The third system, Jazz, is a proprietary system, and hence, was not available. Eclipse JDT is one of the most popular integrated development environment (IDE), which is mainly written in Java. Mozilla mailnews is a popular email client, which is mainly written in C/C++. Tables 1 and 2 provide descriptive statistics for the two studied systems. We used a total of 5,266 bug reports and 26 source code snapshots (as used in Thomas *et al.* [2]).

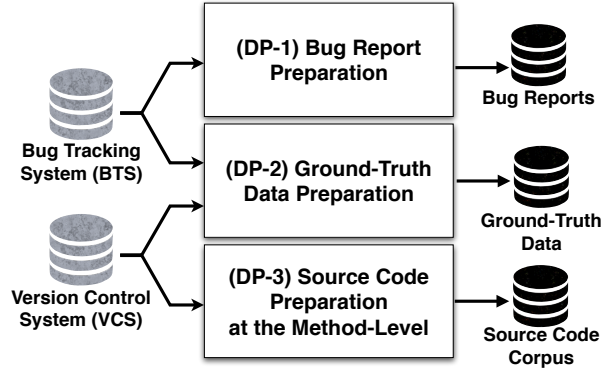

Figure 2: An overview of our data preparation approach (Section 4.2).

### 4.2. Data Preparation

In order to produce the necessary datasets for our study, we first need to prepare the data from the Bug Tracking Systems (BTS) of each studied system. Next, we need to prepare a set of links between bug reports and entities that were changed to resolve the corresponding bugs to create the ground-truth data at the method level. Finally, we need to prepare source code corpus at the method level from the Version Control Systems (VCS) of each studied system. Figure 2 provides an overview of our data preparation approach, which is further divided into the three steps that we describe below.

**(DP-1) Bug Report Preparation.** In IR-based bug localization, we have to collect bug reports, which are treated as a query, and source code entities, which are treated as a document corpus. For each studied system, we obtain the raw bugs data and the source code snapshots at the file-level from Thomas *et al.* [2].

**(DP-2) Ground-Truth Data Preparation.** To create the ground-truth data at the method-level, we need to establish a set of links between bug reports and the methods that were changed to resolve the corresponding bugs. To identify the changed methods, we used the information in the change messages of the Git commit log. Change messages of the Git commit log contains the method names whose content were changed in a given commit.

To identify a bug report in the Git commit log, we used an approach similar to Fischer *et al.* [40], which is also

```
commit b1672031b269719ce6519561e4ea344da64970cb
Author: oliviert <oliviert>
Date:   Tue Nov 3 15:37:46 2009 +0000

    HEAD - Fix for 293777

diff --git a/compiler/org/eclipse/jdt/internal/compiler/lookup/MethodScope.java
          b/compiler/org/eclipse/jdt/internal/compiler/lookup/MethodScope.java
index 4901ba8..00aaffd 100644
--- a/compiler/org/eclipse/jdt/internal/compiler/lookup/MethodScope.java
+++ b/compiler/org/eclipse/jdt/internal/compiler/lookup/MethodScope.java

@@ -469,7 +469,7 @@ public final int recordInitializationStates(FlowInfo flowInfo) {
```

Figure 3: A snippet of the Eclipse JDT commit log.

used in Thomas et al. [2]. This approach parses the Git commit log messages and looks for messages with "fixed" and "bugs" keywords (e.g., "Fixed Bug #293777"). If such a message is found, the approach will establish a link between the change commit and a bug report using the identified bug ID.

To identify the changed methods' names in the change messages, we rely on the change information of the `git log` command. The change information will show the changed methods' names and lines where method declarations occur. Figure 3 shows a snippet of a Git commit log for Eclipse JDT obtained using the command `git log -p`. This commit log provides information about what is modified to resolve bug ID 293777.[1] The line which starts with `@@` in the log shows the name of the method that is changed to fix the bug. As shown in Figure 3, the method `recordInitializationStates` of the file `MethodScope.java` was fixed to resolve the bug ID 293777.

**(DP-3) Source Code Preparation at the Method-level.** To build the source code corpus at the method-level, we use an abstract syntax tree (AST) to extract methods from source code files. In addition to methods, source code files contain source code elements such as attribute and header definitions. Hence, in order not to miss any information which were in the files, we created a dummy method per file. The dummy method for each file contains all statements in the file which do not fall into the body of methods, e.g., attribute definitions. For the Eclipse system, which is written in Java, we use the publicly available JavaParser library.[2] For the Mozilla system, which is written in C++, we use a publicly available C++ method extractor based on regular expressions.[3]

4.3. Classifier Configuration Framework

Table 3 show the summaries of the parameters and the corresponding values that are used in the configuration of the classifiers. The configurations are presented in Thomas et al. [2]. However, for the sake of completeness and clarity, we present them briefly below.

For bug report representation, there are three values: the title of the bug report only (A1); the description only (A2); and both the title and description of the bug report (A3).

For source code entity representation, there are six values. The first three parameters are based on the text of the source code entity itself: the identifier names (B1); comments only (B2); and both identifiers and comments (B3). The other two parameters are based on past bug report (PBR) [7]; using all the PBRs of an entity (B4); and using just the 10 most recent PBRs of an entity (B5). Finally, we consider all possible data for an entity: its identifier, comments, and all PBRs (B6).

There are three common preprocessing steps: splitting identifiers; removing stop words; and stemming using the Porter stemming algorithm. We tested a total of 8 possible preprocessing techniques (C0-C7).

There are two families of classifiers: IR-based classifiers and entity metric-based classifiers. For IR-based classifiers, we consider Vector Space Model (VSM), Latent Semantic Indexing (LSI), and Latent Dirichlet Allocation (LDA) models.

**VSM** [41] is a simple algebraic model based on the term-document matrix of a corpus. The rows of the matrix are represented by unique terms collected from the corpus while the columns represent unique documents. When a term is contained in a document, the intersection of the term row and document column will hold the weight of the term, otherwise zero. The similarity between two documents will increase as the number of common terms they have increases.

The VSM model has two parameters: term weighting and similarity score. For term weighting, we considered the tf-idf (D1) and sublinear tf-idf (D2) weighting schemes, as well as the more basic Boolean (D3) weighting scheme. tf-idf is computed as the product of the number of occurences of a term in a document and inverse of number of documents containing the term. In sublinear tf-idf, the term frequency is replaced with logarithm of the term frequency. The Boolean weighting scheme assigns one if a term occurs in a document and zero, otherwise.

---

[1] https://bugs.eclipse.org/bugs/show_bug.cgi?id=293777
[2] https://code.google.com/p/javaparser/
[3] https://github.com/SAILResearch/replication-ist_bug_localization/

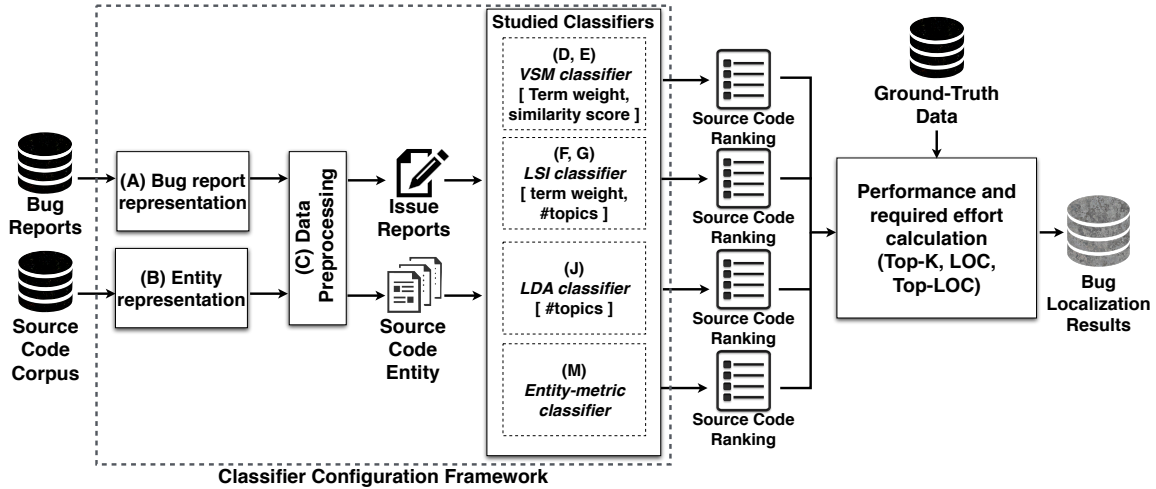

Figure 4: An overview of our data analysis approach that is applied to address our research questions (RQs).

For similarity score, we considered both the cosine (E1) and overlap (E2) similarity scores [42]. Cosine similarity is computed as the ratio of the dot product of two document vectors to the product of their Euclidean length. The overlap similarity is defined as the ratio of the number of common terms of two documents to the size of the smallest document. We use the implementation of VSM that is provided by Apache Lucene.[4]

**LSI** [43] is an extension to VSM which assumes that there is some latent structure in word usage that is partially obscured by variability in word choice. Singular value decomposition (SVD) is used as a means to project the original term-document matrix into three new matrices: a topic-document matrix $D$; a term-topic matrix $T$; and a diagonal matrix $S$ of eigenvalues. The dimension in the projected latent semantic space represents the number of topics which is less than the original matrix. A topic contains list of terms that are related by collocation. Unlike VSM, LSI could consider two documents to be similar even if they have no common term. While computing similarity, LSI sees if the terms in the two documents are from the same topic, rather than term overalp.

The LSI model has three parameters: term weighting, similarity score, and number of topics. We considered the same three weighting schemes as we did for the VSM model (F1-F3). We hold the similarity score constant at cosine (H1). Finally, we considered four values for the number of topics: 32, 64, 128, and 256 (G32-G256). We use the implementation LSI that is provided by Gensim.[5]

**LDA** [44] is a popular statistical topic model which provides a means to automatically index, search, and cluster documents that are unstructured and unlabeled. In LDA, documents are represented as a mixture of words taken from different latent topics, where a topic is characterized by a distribution over words. A topic is defined before any data is generated as a distribution over a fixed vocabulary. LDA rank terms representing a topic using a probability of membership. The membership probability indicates the level of representativeness of the term in the respective topics in which it is found. In LDA, documents are assumed to be generated by randomly choosing a topic from a selected topic distribution and assigning the topic for a given term in a document.

The LDA model has five parameters: number of topics, a document-topic smoothing parameter, a topic-word smoothing parameter, number of sampling iterations, and similarity score. We considered four values for the number of topics: 32, 64, 128, and 256 (J32-J256), to be consistent with the LSI model. Finally, we considered the conditional probability score (N1). We use the implementation of LDA that is provided by MALLET Topic Modeling.[6]

Previous studies used the Entity Metric (EM) to predict buggy source code entities [45] and locate bugs in the source code [2]. EM measures source code features such as line of code and past bug proneness to predict and locate buggy source code entities. EM-based classifiers have only a single parameter, entity metric, which is used to determine the bug-proneness of an entity. We considered four metrics: the Lines of Code (LOC) of an entity;[7] the churn of an entity (i.e., we computed the summation of lines added and deleted from the `git log --numstat` command); the cumulative bug count of an entity; the new bug count of an entity.

---

[4] https://lucene.apache.org/core/

[5] https://radimrehurek.com/gensim/
[6] http://mallet.cs.umass.edu/
[7] https://www.dwheeler.com/sloccount/

Table 3: The configuration parameters and the values of the IR (e.g., VSM, LSI, and LDA) and EM family of classifiers, as proposed by [2].

| Parameter | Value |
|---|---|
| *Parameters common to all IR classifiers* | |
| (A) Bug report representation | A1 (Title only) |
| | A2 (Description only) |
| | A3 (Title + description) |
| (B) Entity representation | B1 (Identifiers only) |
| | B2 (Comments only) |
| | B3 (Idents + comments) |
| | B4 (PBR-All) |
| | B5 (PBR-10 only) |
| | B6 (Idents+comments+PBR-All) |
| (C) Preprocessing steps | C0 (None) |
| | C1 (Split only) |
| | C2 (Stop only) |
| | C3 (Stem only) |
| | C4 (Split + stop) |
| | C5 (Split + stem) |
| | C6 (Stop + stem) |
| | C7 (Split + stop + stem) |
| *Parameters for VSM only* | |
| (D) Term weight | D1 (tf-idf) |
| | D2 (Sublinear tf-idf) |
| | D3 (Boolean) |
| (E) Similarity metric | E1 (Cosine) |
| | E2 (Overlap) |
| *Parameters for LSI only* | |
| (F) Term weight | F1 (tf-idf) |
| | F2 (Sublinear tf-idf) |
| | F3 (Boolean) |
| (G) Number of topics | G32 (32 topics) |
| | G64 (64 topics) |
| | G128 (128 topics) |
| | G256 (256 topics) |
| (H) Similarity metric | H1 (Cosine) |
| *Parameters for LDA only* | |
| (I) Number of iterations | I1 (Until model convergence) |
| (J) Number of topics | J32 (32 topics) |
| | J64 (64 topics) |
| | J128 (128 topics) |
| | J256 (256 topics) |
| (K) $\alpha$ | K1 (Optimized based on $K$) |
| (L) $\beta$ | L1 (Optimized based on $K$) |
| (N) Similarity metric | N1 (Conditional probability) |
| *Parameters for EM only* | |
| (M) Metric | M1 (Lines of code) |
| | M2 (Churn) |
| | M3 (New bug count) |
| | M4 (Cumulative bug count) |

To quantify the performance of all possible classifiers, we used a full factorial design. We explored every possible combination of parameter values. In this study, we have 3,168 IR-based classifiers and 4 entity metric-based classifiers. Thus, we have 3,172 classifiers under test. We run all 3,172 classifiers on the data for the two systems at the method level.

## 5. Case Study Results

In this section, we present our research questions (RQs) and their results. For each research question, we discuss the approach that we followed to answer the RQs.

*(RQ1) Can IR-based classifier configurations significantly impact the top-$k$ performance of method-level bug localization?*

**Approach**. To investigate the impact of classifier configuration for method-level bug localization, we use the framework proposed by Thomas et al. [2]. The framework is summarized in Section 4.3. We executed all 3,172 configurations of bug localization approaches at the method-level (see Figure 4). To compare the different classifier configurations, we computed the top-$k$ performance metric for each configuration. The top-$k$ performance metric is described below.

**Evaluation Metric**. Top-$k$ performance metric is the most frequently-used evaluation metric to assess the performance of IR-based bug localization approaches [5, 6, 2, 7, 3]. IR-based bug localization approaches rank source code entities based on the entities' similarity to a query formulated from the bug report. The entities ranked at the top are considered to be the most relevant to start fixing the reported bug. Developers usually examine the ranked entities in the top $k$, starting sequentially from the top, until they find the relevant entity to fix the reported bug. Top-$k$ performance measures the percentage of bug reports for which at least one relevant source code entity was returned in the top $k$ ranked entities. Formally, the top-$k$ performance of a classifier $C_j$ is

$$\text{top-}k(C_j) = \frac{1}{|Q|} \sum_{i=1}^{|Q|} I(\exists d \in D | rel(d, q_i) \wedge r(d|C_j, q_i) \leq k), \quad (1)$$

where $|Q|$ is the number of queries, $q_i$ is an individual query, $rel(d, q_i)$ returns whether entity $d$ is relevant to query $q_i$, $r(d|C_j, q_i)$ is the rank of $d$ given by $C_j$ in relation to $q_i$, and $I$ is the indicator function, which returns 1 if its argument is true and 0 otherwise. For example, a top-20 performance value of 0.25 indicates that for 25% of the bug reports, at least one relevant source code entity was returned in the top 20 results. Following Thomas et al. [2], we choose the $k$ value to be 20.

**Results**. **The choice of classifier configuration impacts the top-$k$ performance from 0.44% to 36%.** Table 4 presents the top-20 performance value of the best and worst four top-$k$ configurations of method-level bug localization. For Eclipse, the best top-$k$ configuration localizes 1,337 bugs (34.3%), while the worst top-$k$ configuration localizes none of the bugs in the top-20 ranked methods. For Mozilla, the best top-$k$ configuration localizes 514 bugs (37.6%), while the worst top-$k$ configuration localizes 12 bugs (0.88%) in the top-20 ranked methods. The wide top-$k$ performance range indicates that classifier configuration plays an important role in the top-$k$ performance of classifiers. Using inappropriate configuration could result in poor top-$k$ performance.

Table 4: The best four configurations and the worst four configurations for method-level bug localization, for each classifier family (VSM, LSI, LDA, and EM) and each studied system. The configurations are ordered according to their top-20 performance.

| **VSM** | | | **LSI** | | | **LDA** | | | **EM** | | |
|---|---|---|---|---|---|---|---|---|---|---|---|
| Rank | Configuration | Top-20 | Rank | Configuration | Top-20 | Rank | Configuration | Top-20 | Rank | Config. | Top-20 |
| **Eclipse system** | | | | | | | | | | | |
| 1 | A1.B4.C5.D1.E1 | 0.343 | 1 | A1.B4.C7.F1.G256 | 0.307 | 1 | A1.B4.C6.J32.K1 | 0.083 | 1 | M2 | 0.117 |
| 2 | A1.B4.C3.D1.E1 | 0.343 | 2 | A1.B4.C6.F1.G256 | 0.307 | 2 | A1.B4.C6.J64.K1 | 0.081 | 2 | M4 | 0.037 |
| 3 | A1.B4.C7.D1.E1 | 0.340 | 3 | A1.B4.C3.F1.G256 | 0.303 | 3 | A1.B4.C2.J64.K1 | 0.080 | 3 | M1 | 0.027 |
| 4 | A1.B4.C6.D1.E1 | 0.340 | 4 | A1.B4.C5.F1.G256 | 0.302 | 4 | A1.B4.C4.J128.K1 | 0.079 | 4 | M3 | 0.027 |
| 861 | A3.B2.C3.D1.E2 | 0.007 | 1725 | A2.B2.C1.F1.G32 | 0.004 | 574 | A1.B2.C2.J32.K1 | 0.001 | - | | |
| 862 | A2.B2.C0.D2.E2 | 0.006 | 1726 | A2.B2.C1.F3.G32 | 0.004 | 574 | A1.B2.C2.J256.K1 | 0.001 | - | | |
| 863 | A2.B2.C3.D1.E2 | 0.006 | 1727 | A2.B2.C1.F2.G32 | 0.004 | 575 | A1.B2.C0.J256.K1 | 0.001 | - | | |
| 864 | A2.B2.C0.D1.E2 | 0.006 | 1728 | A2.B2.C0.F1.G32 | 0.004 | 576 | A1.B2.C2.J128.K1 | 0.000 | - | | |
| **Mozilla system** | | | | | | | | | | | |
| 1 | A3.B1.C7.D1.E1 | 0.376 | 1 | A3.B3.C5.F2.G256 | 0.308 | 1 | A3.B6.C7.J32.K1 | 0.096 | 1 | M2 | 0.087 |
| 2 | A3.B3.C7.D1.E1 | 0.376 | 2 | A3.B1.C5.F2.G256 | 0.303 | 2 | A3.B6.C4.J32.K1 | 0.091 | 2 | M1 | 0.062 |
| 3 | A3.B1.C5.D1.E1 | 0.373 | 3 | A3.B3.C1.F2.G256 | 0.282 | 3 | A3.B6.C7.J64.K1 | 0.090 | 3 | M3 | 0.033 |
| 4 | A3.B3.C5.D1.E1 | 0.370 | 4 | A3.B1.C2.F3.G256 | 0.278 | 4 | A3.B6.C7.J128.K1 | 0.087 | 4 | M4 | 0.032 |
| 861 | A2.B3.C0.D1.E2 | 0.004 | 1725 | A2.B1.C0.F1.G32 | 0.003 | 574 | A3.B2.C7.J64.K1 | 0.010 | - | | |
| 862 | A3.B3.C0.D1.E2 | 0.003 | 1726 | A2.B2.C2.F3.G32 | 0.003 | 574 | A1.B2.C2.J32.K1 | 0.010 | - | | |
| 863 | A2.B1.C0.D1.E2 | 0.003 | 1727 | A1.B2.C6.F3.G32 | 0.003 | 575 | A1.B2.C2.J128.K1 | 0.010 | - | | |
| 864 | A3.B1.C0.D1.E2 | 0.003 | 1728 | A2.B3.C0.F1.G32 | 0.002 | 576 | A3.B2.C5.J256.K1 | 0.008 | - | | |

Table 5: The likelihood of randomly picking a configuration within 20% of the best top-$k$ performance.

| Classifier | # Config. | 1% | 5% | 10% | 15% | 20% |
|---|---|---|---|---|---|---|
| **Eclipse system** | | | | | | |
| VSM | 864 | 0.005 | 0.010 | 0.019 | 0.023 | 0.042 |
| LSI | 1,728 | 0.001 | 0.002 | 0.010 | 0.020 | 0.034 |
| LDA | 576 | 0.002 | 0.007 | 0.017 | 0.035 | 0.054 |
| EM | 4 | 0 | 0 | 0 | 0 | 0 |
| **Mozilla system** | | | | | | |
| VSM | 864 | 0.003 | 0.007 | 0.014 | 0.041 | 0.065 |
| LSI | 1,728 | 0.001 | 0.001 | 0.002 | 0.010 | 0.020 |
| LDA | 576 | 0.002 | 0.002 | 0.009 | 0.009 | 0.023 |
| EM | 4 | 0 | 0 | 0 | 0 | 0 |

Table 6: Top-20 performance dispersions of method-level bug localization for Eclipse and Mozilla.

| | Min. | $1^{st}$ Qu. | Med | Mean | $3^{rd}Qu.$ | Max | Variance |
|---|---|---|---|---|---|---|---|
| **Eclipse system** | | | | | | | |
| VSM | 0.006 | 0.031 | 0.073 | 0.098 | 0.152 | 0.343 | 0.0073 |
| LSI | 0.004 | 0.023 | 0.054 | 0.089 | 0.153 | 0.308 | 0.0062 |
| LDA | 0.000 | 0.005 | 0.017 | 0.025 | 0.041 | 0.083 | 0.0005 |
| EM | 0.027 | 0.027 | 0.033 | 0.052 | 0.058 | 0.117 | 0.0019 |
| **Mozilla system** | | | | | | | |
| VSM | 0.034 | 0.078 | 0.126 | 0.146 | 0.189 | 0.376 | 0.0067 |
| LSI | 0.025 | 0.093 | 0.148 | 0.138 | 0.177 | 0.308 | 0.0034 |
| LDA | 0.009 | 0.027 | 0.045 | 0.044 | 0.058 | 0.096 | 0.0003 |
| EM | 0.032 | 0.033 | 0.048 | 0.054 | 0.068 | 0.087 | 0.0007 |

**Among the four types of classifiers, VSM achieves the best top-$k$ performance.** When comparing the best top-$k$ configuration of each classifier, on average, VSM is 1.2 to 4 times better than other classifiers (i.e., LSI, LDA, and EM). We suspect that VSM outperforms others has to do with the similar textual characteristics between bug reports and source codes. To see if increasing the number of topics changes the results, we run LSI and LDA using 512 topics and found the result to still hold. We suspect that increasing the number of topics will produce more granular topics, instead of totally different topics [46]. Thus, increasing the number of topics has little impact on the top-k performance.

**The likelihood of randomly picking a configuration that performs within 20% of the best top-$k$ classifier (VSM) configuration is on average 5.4%.** Table 5 shows the likelihood of configurations that perform within 1, 5, 10, 15, and 20 percent of the best top-$k$ performing classifier configuration of each classifier. The second column in the tables indicates the total number of configurations for the respective classifiers. EM has only four configurations, hence, we did not compute the likelihood of randomly picking the best top-$k$ configuration. While using the best top-$k$ classifier (VSM), the likelihood of randomly picking a configuration that performs within

20% of the best top-$k$ configuration is on average 5.4% for method level. The low likelihood indicates that there are only few classifier configurations that perform close to the best performing configuration. The statistical summaries of classifier configurations performance also indicate that the performance of the majority of classifier configurations is low as compared to the best top-$k$ performing classifier (see Table 6). Hence, finding the best top-$k$ configuration is difficult.

> *Summary:* Configuration has a large impact on the top-$k$ performance of method-level bug localization, suggesting that using an inappropriate configuration could result in poor top-$k$ performance. There are only few classifier configurations that perform close to the best performing configuration, indicating that finding the best top-$k$ configuration is difficult. Hence, practitioners are in need of guidance to help them in selecting the optimal configurations.

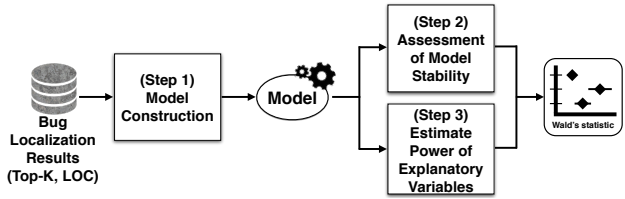

Figure 5: An overview of our sensitivity analysis approach.

<u>Parameter Sensitivity Analysis</u>. Our results indicate that locating the best configuration is difficult. Hence, we would like to explore which of the parameters has the most impact on the performance of a classifier. To perform such sensitivity analysis, we build regression models. The models help to understand the relationship between classifier configurations and the top-$k$ performance of a classifier. To study the importance of each configuration parameter and examine the relative contribution (in terms of explanatory power) of each configuration parameter to the regression model, we perform an ANOVA analysis [47, 48]. As suggested by our recent work, we use ANOVA Type II [49, 50]. Figure 5 shows an overview of our sensitivity analysis approach. We describe each step of our approach below.

**(Step-1) Model Construction.** We build regression models to explain the relationship that classifier configurations have on the top-$k$ performance of a classifier. A regression model fits a line of the form $y = \beta_0 + \beta_1 x_1 + \beta_2 x_2 + ... + \beta_n x_n$ to the data, where $y$ is the dependent variable and each $x_i$ is an explanatory variable. In our models, the dependent variable is the top-$k$ performance and the explanatory variables are the set of parameters outlined in Table 3. We fit our regression models using the Ordinary Least Squares (OLS) technique using the `ols` function provided by the `rms` package [51], as suggested by recent studies [52, 53, 54].

**(Step-2) Assessment of Model Stability.** We evaluate the fit of our models using the *Adjusted* $R^2$, which provides a measure of fit that penalizes the use of additional degrees of freedom. However, since the adjusted $R^2$ is measured using the same data that was used to train the model, it is inherently upwardly biased, i.e., "optimistic". We estimate the optimism of our models using the following bootstrap-derived approach [48, 55].

First, we build a model from a bootstrap sample, i.e., a dataset sampled with replacement from the original dataset, which has the same population size as the original dataset. Then, the optimism is estimated using the difference of the adjusted $R^2$ of the bootstrap model when applied to the original dataset and the bootstrap sample. Finally, the calculation is repeated 1,000 times in order to calculate the average optimism. This average optimism is subtracted from the adjusted $R^2$ of the model fit on the original data to obtain the optimism-reduced adjusted $R^2$. The smaller the average optimism, the higher the stability of the original model fit.

**(Step-3) Estimate Power of Explanatory Variables.** We perform an ANOVA analysis of each classifier configuration using the Wald $\chi^2$ maximum likelihood (a.k.a., "chunk") test. The larger the Wald $\chi^2$ value, the larger the impact that a particular explanatory variable has on the response [48].

Finally, we present both the raw Wald $\chi^2$ values, and its bootstrap 95 percentile confidence interval. High Wald $\chi^2$ indicates high impact on the top-$k$ performance of classifier.

**For VSM, the choice of both term weight and source code representation consistently appears as the most important parameters at the method level.** Figure 6 shows that the choice of both *term weight* and *source code representation* (i.e., B and D) has the highest explanatory power for VSM only at the method level.

**For LSI and LDA, while the choice of data representation and data preprocessing consistently appears as the most important parameters, the *number of topics* in LSI and LDA consistently appear as the least important parameters.** Figure 6 shows that the choice of *source code representation* (i.e., B) has the highest explanatory power. Conversely, the number of topics for LSI and LDA (i.e., G and J) have a low sensitivity. Hence, the number of topics has minimal impact on the performance of classifiers. Our results also confirm the finding of Bigger *et al.* [56]. Our findings suggest that practitioners should carefully select configurations related to the source code representation and not to be as concerned about the number of topics.

> *Summary:* Configurations related to the source code representation has the most impact on the top-$k$ performance, suggesting that practitioners should carefully select the source code representation settings and not to be as concerned about the number of topics.

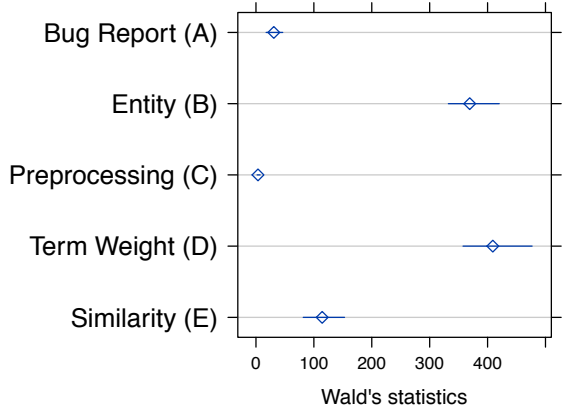

(a) Eclipse (VSM)

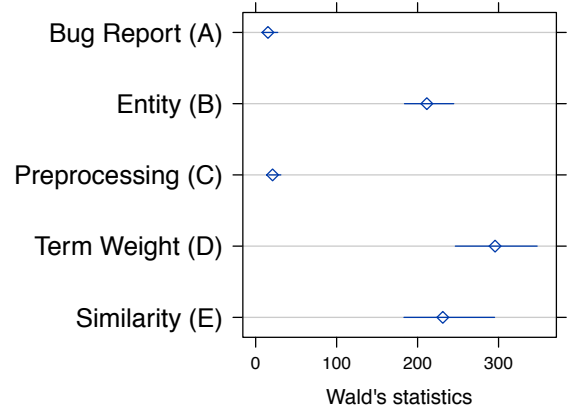

(b) Mozilla (VSM)

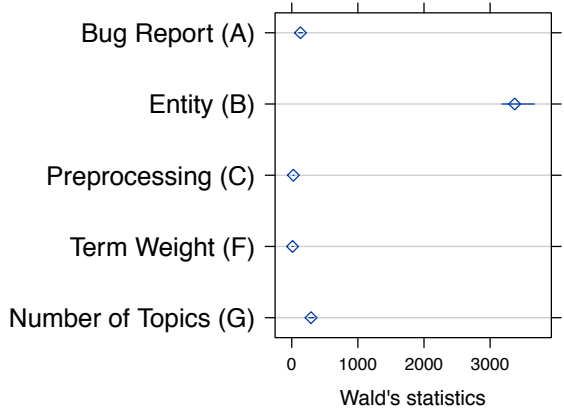

(c) Eclipse (LSI)

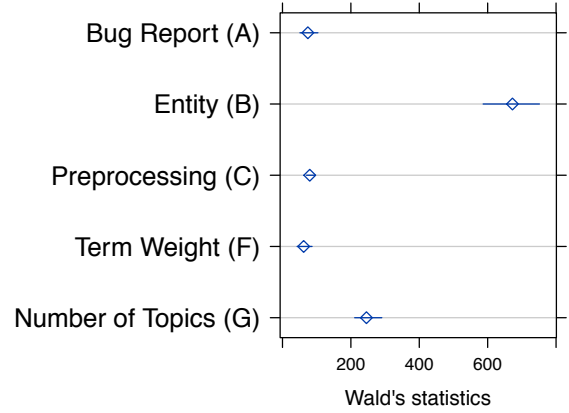

(d) Mozilla (LSI)

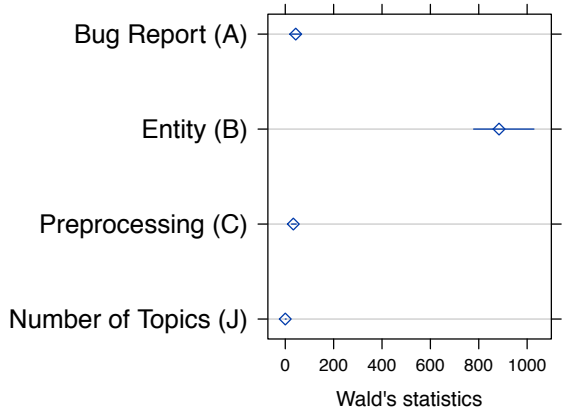

(e) Eclipse (LDA)

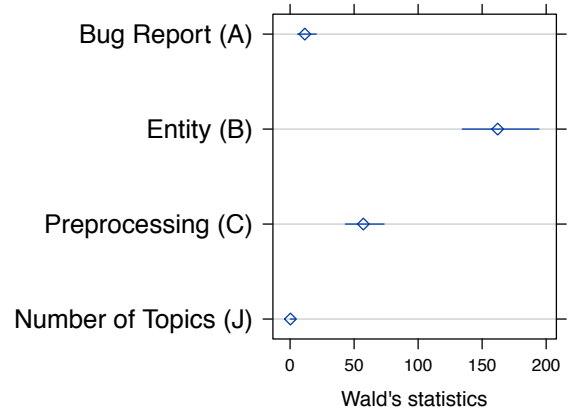

(f) Mozilla (LDA)

Figure 6: Distribution of Wald's statistics for each configuration parameter. A horizontal blue line indicates the bootstrap percentile 95% confidence interval, where a diamond shape indicates the median Wald's statistic for each configuration parameter of each studied system *at the method level*. High Wald's statistics indicates high impact on the top-$k$ performance of the classifier.

*RQ2: Can IR-based classifier configurations significantly impact the required effort for method-level bug localization?*

**Approach**. To address this research question, we used lines of code (LOC) as a proxy to measure the *required effort* to locate bugs. Arisholm *et al.* [31] point out that the cost of source code quality assurance activities, such as code inspection, are roughly proportional to the size of source code entity. For every classifier configuration, we computed the cumulative sum of LOC a developer would need to read to find the first relevant entity to a bug report. The effort is computed for all configurations that are ranked based on top-$k$ performance at the method-level (see Table 4). The results are presented using horizontal boxplots. In order to better see the distribution of the efforts required to locate the first buggy entities, we used 50,000 LOC as the maximum LOC to be examined.

**Results**. **The required effort of the classifier configurations varies from 4,395 to 50,000 LOC.** Table 7 presents horizontal boxplots of the cumulative sum of LOC that a developer would read to locate the first buggy entity. The boxplots are shown for the best four and the worst four top-$k$ configurations at the method granularity level. The average median required effort for the best top-$k$ configurations is 4,395 LOC, while the required effort for the worst top-$k$ configuration is 50,000 LOC, respectively. The observed wide range of required effort between the best and worst top-$k$ configuration indicates that the choice of configuration also impacts effort.

Classifier configurations which give similar top-$k$ performance could require drastically different amount of efforts. While evaluating classifier configurations using top-$k$ performance metrics, two configurations could achieve similar performance. However, we find that the required effort to locate bugs using the two configurations could be a median difference of 2.6 times and vary up to 15 times. For example, for Eclipse at the method-level, the second best LDA configuration (`A1.B4.C6.J64.K1`) and the third best LDA configuration (`A1.B4.C2.J64.K1`) achieve a top-$k$ performance of 0.081 and 0.080, respectively. However, the median required effort by the two configurations is 13,820 LOC and 50,000 LOC, respectively. Hence, we recommend that researchers should also take into consideration the required effort to locate bugs while comparing the performance of classifier configurations using top-$k$ metrics. We also computed the Spearman rank correlation [57, 58] between the rank of the configurations ordered by top-$k$ performance and the effort for method granularity level. We find that the Spearman rank correlation values for Eclipse and Mozilla are 0.9 and 0.8, respectively. The results indicate that the configurations that perform well at top-$k$ performance tend to give the lowest effort.

**Among the four types of classifiers, VSM requires the least effort.** For Eclipse and Mozilla, Table 8 shows that the least efforts are 2,025 LOC and 3,090 LOC, while the effort required by the best VSM top-$k$ configurations (`A1.B4.C5.D1.E1` and `A3.B1.C7.D1.E1`) are 2,602 LOC and 6,188 LOC, respectively. Hence, our findings suggest that the best VSM top-$k$ configuration requires the least effort to locate bugs at the method-level.

**The likelihood of randomly picking a configuration that performs within 20% of the least effort is on average 1%.** Table 9 shows the likelihood of configurations that perform within 1, 5, 10, 15, and 20 percent of the classifier configuration that requires the least effort of each classifier. The second column in the tables indicates the total number of configurations for the respective classifiers. EM has only four configurations, hence, we did not compute the likelihood of randomly picking the configuration that requires the least effort. While using the classifier that requires the least effort (VSM, LSI), the likelihood of randomly picking a configuration that performs within 20% of the least effort configuration is on average 1% for method level, respectively. The low likelihood indicates that there are only a few classifier configurations that require effort close to the least effort configuration. The statistical summaries of the required effort for different classifier configurations also indicate that the effort of the majority of classifier configurations is large as compared to the least effort classifier (see Table 8). Hence, this indicates that finding the least effort configuration is difficult and more difficult than finding the best top-$k$ configuration.

> *Summary: Configuration has a large impact on the required effort of bug localization at the method level, suggesting that practitioners should take into consideration required effort to locate bugs while comparing the performance of classifier configurations using top-$k$ metrics. There are only few classifier configurations that perform close to the configuration that requires the least effort, indicating that finding the least effort configuration is difficult. Hence, practitioners are in need of guidance to help them in selecting the optimal configurations.*

**Parameter Sensitivity Analysis**. We again use the high-level approach of Figure 5 to understand the relationship that classifier configurations have on the required effort.

**In contrast to the top-$k$ performance, the most important parameters are inconsistent across classifiers and systems for the method level classifiers.** Figure 6 shows that there is no consistency in the most important parameters across classifiers and systems for the method level, indicating that the parameters are more susceptible to the required effort. Such inconsistency suggests that the impact that parameter configurations has on the required effort is larger than their impact that has on the top-$k$ performance. Conversely, the number of topics in LSI and LDA tends to appear as the least important parameters, which is consistent to the top-$k$ performance.

Table 7: The amount of lines of code (Median LOC) a developer needs to read in order to locate the first buggy entity of the best four configurations and the worst four configuration of bug localization at method-level, for each classifier family (VSM, LSI, LDA, and EM) and each studied system. The configurations are ordered according to their top-20 performance.

| | VSM | | | LSI | | | LDA | | | EM | |
|---|---|---|---|---|---|---|---|---|---|---|---|
| Rank | Configuration | Median | Rank | Configuration | Median | Rank | Configuration | Median | Rank | Configuration | Median |
| **Eclipse system at the method-level** | | | | | | | | | | | |
| 1 | A1.B4.C5.D1.E1 | 2,602 | 1 | A1.B4.C7.F1.G256 | 2,399 | 1 | A1.B4.C6.J32.K1 | 15,190 | 1 | M2 | 9,230 |
| 2 | A1.B4.C3.D1.E1 | 2,602 | 2 | A1.B4.C6.F1.G256 | 2,412 | 2 | A1.B4.C6.J64.K1 | 13,820 | 2 | M4 | 12,580 |
| 3 | A1.B4.C7.D1.E1 | 2,639 | 3 | A1.B4.C3.F1.G256 | 2,719 | 3 | A1.B4.C2.J64.K1 | 50,000 | 3 | M1 | 50,000 |
| 4 | A1.B4.C6.D1.E1 | 2,639 | 4 | A1.B4.C5.F1.G256 | 2,711 | 4 | A1.B4.C4.J128.K1 | 50,000 | 4 | M3 | 15,660 |
| 861 | A3.B2.C3.D1.E2 | 50,000 | 1725 | A2.B2.C1.F1.G32 | 50,000 | 574 | A1.B2.C2.J32.K1 | 50,000 | - | | |
| 862 | A2.B2.C0.D2.E2 | 50,000 | 1726 | A2.B2.C1.F3.G32 | 50,000 | 574 | A1.B2.C2.J256.K1 | 50,000 | - | | |
| 863 | A2.B2.C3.D1.E2 | 50,000 | 1727 | A2.B2.C1.F2.G32 | 50,000 | 575 | A1.B2.C0.J256.K1 | 50,000 | - | | |
| 864 | A2.B2.C0.D1.E2 | 50,000 | 1728 | A2.B2.C0.F1.G32 | 50,000 | 576 | A1.B2.C2.J128.K1 | 50,000 | - | | |
| **Mozilla system at the method-level** | | | | | | | | | | | |
| 1 | A3.B1.C7.D1.E1 | 6,188 | 1 | A3.B3.C5.F2.G256 | 8,614 | 1 | A3.B6.C7.J32.K1 | 24,360 | 1 | M2 | 20,760 |
| 2 | A3.B3.C7.D1.E1 | 5,932 | 2 | A3.B1.C5.F2.G256 | 9,127 | 2 | A3.B6.C4.J32.K1 | 29,350 | 2 | M1 | 50,000 |
| 3 | A3.B1.C5.D1.E1 | 8,744 | 3 | A3.B3.C1.F2.G256 | 11,270 | 3 | A3.B6.C7.J64.K1 | 24,650 | 3 | M3 | 31,080 |
| 4 | A3.B3.C5.D1.E1 | 8,453 | 4 | A3.B1.C2.F3.G256 | 11,810 | 4 | A3.B6.C7.J128.K1 | 25,640 | 4 | M4 | 34,430 |
| 861 | A2.B3.C0.D1.E2 | 50,000 | 1725 | A2.B1.C0.F1.G32 | 50,000 | 574 | A3.B2.C7.J64.K1 | 50,000 | - | | |
| 862 | A3.B3.C0.D1.E2 | 50,000 | 1726 | A1.B2.C2.F3.G32 | 50,000 | 574 | A1.B2.C2.J32.K1 | 50,000 | - | | |
| 863 | A2.B1.C0.D1.E2 | 50,000 | 1727 | A1.B2.C6.F3.G32 | 50,000 | 575 | A1.B2.C2.J128.K1 | 50,000 | - | | |
| 864 | A3.B1.C0.D1.E2 | 50,000 | 1728 | A2.B3.C0.F1.G32 | 50,000 | 576 | A3.B2.C5.J256.K1 | 50,000 | - | | |

Table 8: The median required effort dispersion amongst classifier families.

| | Min. | $1^{st}$ Qu. | Med | Mean | $3^{rd}Qu.$ | Max | Var |
|---|---|---|---|---|---|---|---|
| **Eclipse system** | | | | | | | |
| VSM | 2,400 | 6,788 | 12,830 | 20,320 | 31,680 | 50,000 | 293 |
| LSI | 2,025 | 5,440 | 12,990 | 21,460 | 43,410 | 50,000 | 341 |
| LDA | 11,920 | 16,330 | 50,000 | 35,080 | 50,000 | 50,000 | 262 |
| EM | 9,230 | 11,740 | 14,120 | 21,870 | 24,250 | 50,000 | 358 |
| **Mozilla system** | | | | | | | |
| VSM | 3,090 | 9,439 | 36,590 | 30,530 | 50,000 | 50,000 | 357 |
| LSI | 3,709 | 15,180 | 20,780 | 25,370 | 33,680 | 50,000 | 202 |
| LDA | 24,360 | 50,000 | 50,000 | 47,890 | 50,000 | 50,000 | 28 |
| EM | 20,760 | 28,500 | 32,760 | 34,070 | 38,330 | 50,000 | 146 |

Table 9: The likelihood of randomly picking a configuration within 20% of the least effort configuration for method-level bug localization.

| Classifier | # Config. | 1% | 5% | 10% | 15% | 20% |
|---|---|---|---|---|---|---|
| **Eclipse system** | | | | | | |
| VSM | 864 | 0.002 | 0.002 | 0.009 | 0.021 | 0.028 |
| LSI | 1,728 | 0.001 | 0.001 | 0.001 | 0.002 | 0.003 |
| LDA | 576 | 0.003 | 0.003 | 0.007 | 0.010 | 0.016 |
| EM | 4 | 0 | 0 | 0 | 0 | 0 |
| **Mozilla system** | | | | | | |
| VSM | 864 | 0.001 | 0.002 | 0.003 | 0.005 | 0.007 |
| LSI | 1,728 | 0.001 | 0.001 | 0.003 | 0.004 | 0.004 |
| LDA | 576 | 0.002 | 0.003 | 0.003 | 0.005 | 0.005 |
| EM | 4 | 0 | 0 | 0 | 0 | 0 |

> *Summary:* The most important parameters inconsistently appear across classifiers and systems for the method level, indicating that the parameters are more susceptible to the required effort. Conversely, the number of topics in LSI and LDA consistently tends to appear as the least important parameters, indicating that practitioners should not be concerned about the number of topics as compared to other parameters.

*RQ3: Is the most efficient classifier configuration for method-level bug localization also the most efficient configuration for file-level bug localization (and vice versa)?*

**Approach.** Given a list of source code entities ranked by the similarity scores, a developer is expected to focus on resources at the beginning of the list as much as possible. The question to answer in this context is: what is the percentage of bug reports that can be successfully localized, when reviewing only top LOC of the entities ranking. Hence, we extend the top-$k$ performance metric to top-$k_{LOC}$ performance. Top-$k_{LOC}$ performance is the percentage of bug reports for which at least one relevant source code entity is found in the top ranked entities with cumulative sum of executable LOC below $k$. Formally, the top-$k_{LOC}$ performance of a classifier $C_j$ is

$$\text{top-}k_{LOC}(C_j) = \frac{1}{|Q|} \sum_{i=1}^{|Q|} I(\exists d \in D \mid rel(d, q_i) \land s(r(d|C_j, q_i)) \leq k),$$

where $|Q|$ is the number of queries, $q_i$ is an individual query, $rel(d, q_i)$ returns whether entity $d$ is relevant to query $q_i$, $r(d|C_j, q_i)$ is the rank of $d$ given by $C_j$ in relation to $q_i$, $s$ is the cumulative sum of executable LOC of all entities whose rank is less than or equal to $r(d|C_j, q_i)$, and $I$ is the indicator function, which returns 1 if its argument is true and 0 otherwise. For example, a top-10,000$_{LOC}$ performance value of 0.25 indicates that for 25% of the bug reports, at least one relevant source code entity was returned in the top ranked entities whose cumulative sum of LOC is below 10,000. We choose the $k$ value to be 10,000 LOC.

<u>**Evaluation Metric:**</u> To compare the top-$k_{LOC}$ perfor-

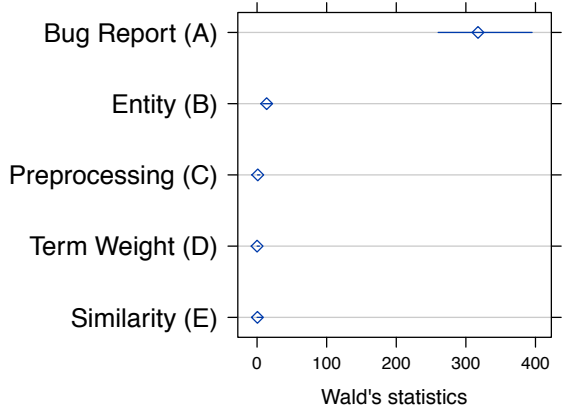
(a) Eclipse (VSM)

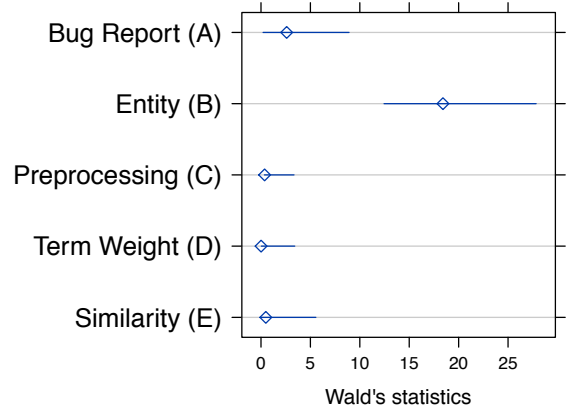
(b) Mozilla (VSM)

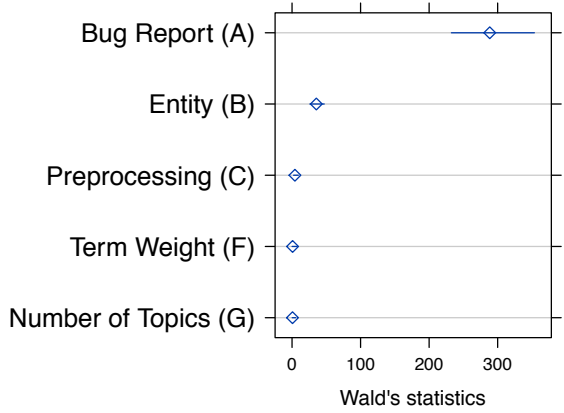
(c) Eclipse (LSI)

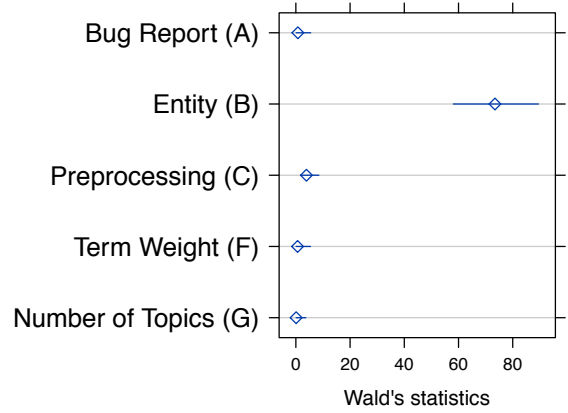
(d) Mozilla (LSI)

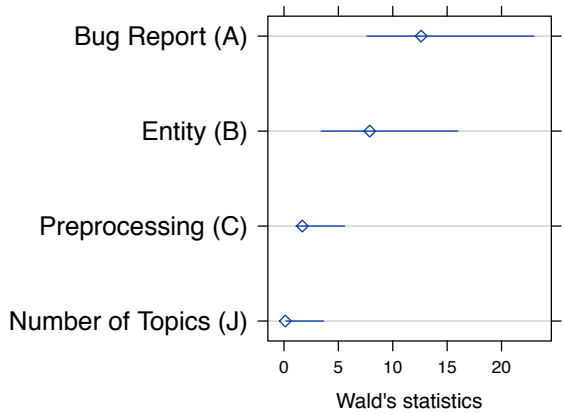
(e) Eclipse (LDA)

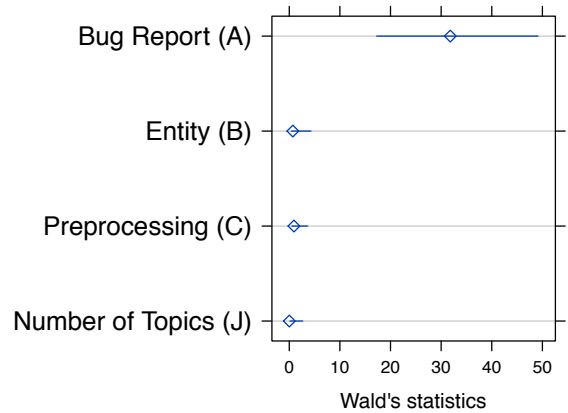
(f) Mozilla (LDA)

Figure 7: Distribution of Wald's statistics for each configuration parameter. A horizontal blue line indicates the bootstrap percentile 95% confidence interval, where a diamond shape indicates the median Wald's statistic for each configuration parameter of each studied system *at the method level*. High Wald's statistics indicates high impact on the required effort of the classifier.

mance of the most efficient classifier configurations, we used the differences between the lift charts of top-$k_{LOC}$ performances obtained at different $k$ values. The comparison is between the top-$k_{LOC}$ performance lift chart of the best configuration of file-level on the method-level with the best configuration of method-level (and vice versa).

**Result**. **The most efficient configuration is an IR-based classifier that uses the Vector Space Model (`A3.B6.C7.D2.E1`)**, with the index built using sub-linear tf-idf term weighting on all available data in the source code entities (i.e., identifiers, comments, and past bug reports for each entity), which has been stopped, stemmed, and split, and queried with all available data in the bug report (i.e., title and description) with cosine similarity.

**The most efficient configuration of file-level bug localization performs close enough to the most efficient configuration of method-level, when used at the method-level.** We used the most efficient file-level configuration at the method-level (*line with diamond*) and compared the results with the results of the most efficient configuration of method-level bug localization (*line with circle*) (see Figure 8). The top-$k_{LOC}$ performance of the most efficient method-level configuration (*line with circle*) is same or higher than the top-$k_{LOC}$ performance of the most efficient file-level configuration (*line with diamond*) while used to locate bugs at the method-level. For Eclipse, the maximum difference where the *line with circle* is higher than the *line with diamond* is 0.004, while for Mozilla, the maximum difference is 0.045. The small differences indicate that the most efficient file-level configuration is also efficient at the method-level.

**The most efficient configuration of method-level bug localization performs close enough to the most efficient configuration of file-level, when used at the file-level.** We compared the most efficient configuration of method-level bug localization at the file-level (*line with square*) with the most efficient configuration of file-level bug localization (*line with triangle*) (see Figure 8). For every $k \leq 10{,}000$, we computed the difference between the top-$k_{LOC}$ performance of the *line with triangle* and the *line with square*. The maximum observed difference for Eclipse is 0.016, while for Mozilla the maximum difference is 0.032. The results of the best efficient configuration comparisons indicate that the best efficient configuration irrespective of the granularity level gives similar result.

> *Summary:* The most efficient classifier configuration obtained at the method-level can also be used at the file-level (and vice versa) without a significant loss of top-$k_{LOC}$ performance.

## 6. Threats to Validity

The key goal of our study is to explore the impact of parameter setting in prior bug localization studies. Hence, our study has much of our threats to validity of prior studies in literature on this topic (especially for internal validity concerns). We do note that a key contribution of our study is the used methodology to quantify and analyze the impact of settings. Such methodology holds independent of threats to validity that arise due to data characteristics.

*6.1. Threats to Internal Validity*

The main internal threat to validity of our study and other bug localization studies in literature lies on the Fischer *et al.* [40]'s technique that we used to collect ground-truth data. Although this technique is commonly used for linking bug reports to source code entities, a number of potential biases can affect the validity of the ground-truth data of our study.

First, wrongly classified bug reports might be included in our datasets, which can impact the top-$k$ performance of bug localization. Herzig *et al.* [59] find that 43% of bug reports are wrongly classified, suggesting that these misclassified bug reports should be excluded from the various analyses. However, an extensive analysis by Kochhar *et al.* [60] finds that wrongly classified bug reports have a negligible impact on the performance of bug localization. Hence, we suspect that this bias does not pose a great threat in our study.

Second, incomplete ground-truth entities can impact the top-$k$ performance of bug localization, another threat that is shared with prior studies. Bird *et al.* [61] find that there are several bug reports that are not identified in the commit logs. To mitigate this threat, such analysis would require deep domain knowledge of the system and a thorough working understanding of how the code is organized and run. Unfortunately, we do not possess this knowledge or understanding.

Third, incorrect ground-truth entities can impact the top-$k$ performance of bug localization. This internal threat to validity is also shared with prior studies. Murphy *et al.* [62, 63] find that developers often refactored source code while fixing bugs. Kawrykow *et al.* [64] find that developers often included unrelated changes (e.g., refactoring, comments modification, code indentation) to the bug-fixing commit. Such unrelated changes are considered as non-buggy entities, which should be excluded from the ground truth data as they do not contain the bug. However, an extensive analysis by Kochhar *et al.* [60] finds that such incorrect ground-truth entities does not have an impact on the performance of bug localization. Hence, we suspect that this bias does not pose a great threat in our study.

Finally, Kochhar *et al.* [60] point out that already localized bug reports (i.e., bug reports where their textual descriptions have already specified the files that contain the bug) can inflate the top-k performance if they are not removed. To ensure if already localized bug reports are affecting the conclusions of our paper, we analyze search the file extensions (e.g., .java, .cpp, .h) using a regular expression for all of the studied bug reports. After we removed these localized bug reports, we find that they do not alter our conclusions. Thus, we suspect that already localized bug reports do not pose a great threat to validity of our conclusions.

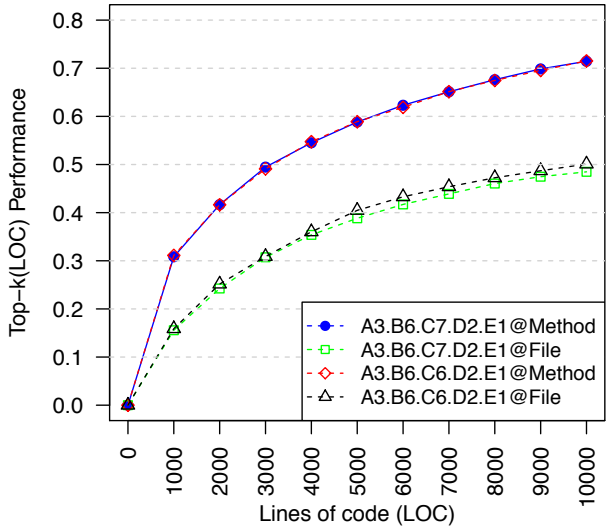 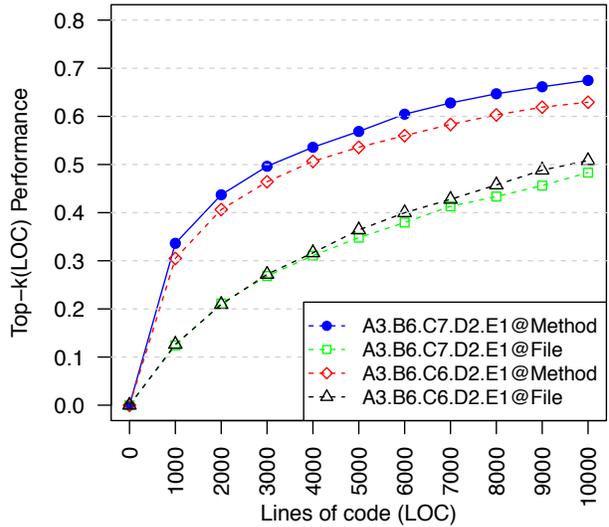

Figure 8: The top-$k_{LOC}$ performance for the most efficient configuration of method-level bug localization used at method-level (*line with circle*) and file-level (*line with square*); and, the best configuration of file-level bug localization used at method-level (*line with diamond*) and file-level (*line with triangle*). The figure shows that the most efficient configuration of file-level bug localization performs close enough to the most efficient configuration of method-level, when used at the method-level (for $k <= 10,000$ LOC).

### 6.2. Threats to External Validity

Threats to external validity are concerned with the generalization of our findings. In this study, we used two large open source software systems, Eclipse and Mozilla, which are large real-world software systems and frequently used in top venue. The data for the systems is obtained from Thomas *et al.* [2] for a fair comparison of RQ1 to file-level bug localization. Thus, our conclusions may not generalize to other software systems. Moreover, we only consider bug reports during a period of 2002-2009 for Eclipse JDT and a period of 2002-2006 for Mozilla. Thus, our conclusions may not generalize to other periods. However, the key goal of our work is to highlight that there is an impact for some datasets. We can add additional datasets but realistically we will never have enough datasets to establish a wide ranging empirical law. The key message to take home from our analyses is that the parameter settings matter. Given the simplicity of considering these settings, we suggest that all future studies explore them (as it might be the case that they matter for their data set). If they do not matter for a particular system, then the wasted efforts are minimal.

### 6.3. Threats to Construct Validity

The main threat to construct validity of our study is related to the effort-based evaluation which is used to analyze method-level bug localization. In our study, we used LOC as a proxy to measure effort as done in prior studies [30, 31, 32]. However, LOC may not reflect actual effort required to assess a code entity. A code entity could be complex and also depend on other entities. Future studies should incorporate factors related to complexity and dependency of the source code entity into the effort-based metrics. This proposed direction for future studies holds for our work and other works in the area of effort-aware bug prediction.

The results of our RQ1 rely on the top-20 performance in order to establish a fair comparison with Thomas *et al.* [2]. Recently, a survey on 386 practitioners by Kochhar *et al.* [65] shows that more than 73.58% of respondents suggested that the top-5 performance is the optimal inspection threshold for bug localization. Thus, future studies should consider top-5 performance as suggested by practitioners. Since the top-$k$ performance does not measure the number of relevant buggy entities in the top-$k$ ranked list, other performance measures like Mean Reciprocal Rank (MRR) and Recall should be considered in future work. Mean Reciprocal Rank (MRR) is a statistical measure for evaluating any process that produces a list of possible responses to a query. The reciprocal rank of a query response is the multiplicative inverse of the rank of the first correct answer. Recall measures the percentage of the relevant buggy entities that successfully retrieved for a bug report. Nevertheless, the results of our RQ2 show that classifier configurations which give similar top-k performance could require drastically different amount of efforts. Thus, traditional IR evaluation metrics (e.g., top-$k$ performance, MRR, and Recall) should also consider the required effort to locate bugs (e.g., the top-$k_{LOC}$ performance in RQ3) into consideration.

## 7. Conclusion

Previous study showed that classifier configuration has an impact on file-level bug localization. Several bug localization studies, however, are also conducted at the method-level. In this paper, we investigate the impact that the choice of IR-based classifier configuration has on the top-$k$ performance and the required effort to examine source code entities (e.g., files and methods) before locating a bug at the method level. Moreover, we also analyze the classifier sensitivity to parameter value changes. In total, we explore a large space of classifier configurations, 3,172 configurations. Through a case study of 5,266 bug reports of two large-scale software systems (i.e., Eclipse and Mozilla), we make the following observations:

- The choice of classifier configuration impacts the top-$k$ performance from 0.44% to 36% and the required effort from 4,395 to 50,000 LOC, suggesting that using inappropriate configurations could result in poor top-$k$ performance and wasted effort.

- Classifier configurations which give similar top-$k$ performance could require different efforts, suggesting that practitioners should take into consideration required effort to locate bugs while comparing the performance of classifier configurations.

- VSM achieves both the best top-$k$ performance and the least required effort for method-level bug localization.

- The likelihood of randomly picking a configuration that performs within 20% of the best top-$k$ classifier configuration is on average 5.4% and that of the least effort is on average 1%.

- Configurations related to the entity representation of the analyzed data have the most impact on the top-$k$ performance and the required effort, suggesting that practitioners would benefit of guidance on which configuration parameters matter the most.

- The most efficient classifier configuration obtained at the method-level can also be used at the file-level (and vice versa) without a significant loss of top-$k_{LOC}$ performance.

Furthermore, we also repeat our analysis at the file level to extend the findings of [2]. Table 10 summarizes the key findings of the method-level and file-level bug localizations. Our results lead us to conclude that configurations have a large impact on both the top-$k$ performance and the required effort for method-level and file-level bug localization, suggesting that the IR-based configuration settings should be carefully selected and the required effort measure should be included in future bug localization studies.

At the end, we suggest that the most efficient classifier configuration for bug localization is `A3.B6.C7.D2.E1@Method` the Vector Space Model, with the index built using sub-linear tf-idf term weighting on methods (i.e., identifiers, comments, and past bug reports for each entity), which has been stopped, stemmed, and splitted, and queried with all available data in the bug report (i.e., title and description) with cosine similarity. We provide our datasets online in order to encourage future research in the area of IR-based bug localization.[8]

## Acknowledgments

We would like to thank Stephen W. Thomas for giving us access to his classifier configuration framework and ground-truth dataset.

---

[8] https://github.com/SAILResearch/replication-ist_bug_localization/

Table 10: Summary of the main findings. The two cells with gray background indicate the main contributions of Thomas *et al.* [2], while the remaining 18 cells with white background are the main contributions of this paper.

| | File-Level | Method-Level |
|---|---|---|
| Does classifier configuration impact top-$k$ performance of classifiers? | Yes | Yes |
| What is the best top-$k$ performing classifier? | VSM | VSM |
| What is the range between the best and worst top-$k$ performing classifiers? | The top-$k$ performance range is between 0.37% and 67.48%, on average. | The top-$k$ performance range is between 0.44% and 36%, on average. |
| What is the likelihood of randomly picking a configuration that performs within 20% of the best top-$k$ performing classifier? | There is an average of 12.2% chance of randomly picking a configuration that performs within 20% of the best top-$k$ performing classifier. | There is an average 5.4% chance of randomly picking a configuration that performs within 20% of the best top-$k$ performing classifier. |
| What is the likelihood of randomly picking a configuration that performs within 20% of the classifier that requires the least effort? | There is an average of 1% chance of randomly picking a configuration that performs within 20% the classifier that requires the least effort. | There is an average of 2% chance of randomly picking a configuration that performs within 20% the classifier that requires the least effort. |
| Which parameter has the most impact on the performance? (i.e., sensitivity) | Term weight for VSM. Entity representation for LSI and LDA. | Term weight for VSM. Entity representation for LSI and LDA. |
| Which parameter has the most impact on the required effort? (i.e., sensitivity) | Inconsistent results for the most important parameters. However, the number of topics appears as the least important parameters. | Inconsistent results for the most important parameters. However, the number of topics appears as the least important parameters. |
| Does the best top-$k$ performing classifier require the least effort? | VSM is the most top-k performing classifier. However, LSI requires the least effort. VSM requires as much as 2.56 times more effort than LSI. | VSM is the best top-k performing classifier and also requires the least effort. |
| Which IR classifier gives the least effort? | LSI | VSM |
| Is the most efficient classifier configuration for method-level bug localization also efficient for file-level bug localization (and vice versa)? | Yes | Yes |

19